\documentclass[conference]{IEEEtran}
\IEEEoverridecommandlockouts
\usepackage{cite}
\usepackage{amsmath,amssymb,amsfonts}
\usepackage{algorithmic}
\usepackage{graphicx}
\usepackage{textcomp}
\usepackage[utf8]{inputenc}

\usepackage{xcolor}
\usepackage{hyperref}
\usepackage{orcidlink}
\usepackage{todonotes}
\usepackage{listings}
\usepackage{cleveref}
\usepackage[most]{tcolorbox}
\usepackage[inline]{enumitem}

\usepackage{inconsolata}  

\definecolor{codebg}{RGB}{250,250,250}
\definecolor{keywordblue}{rgb}{0.26, 0.35, 0.75}
\definecolor{commentgray}{gray}{0.45}
\definecolor{stringred}{rgb}{0.65,0.13,0.13}
\definecolor{bordergray}{gray}{0.6}
\definecolor{errorred}{RGB}{210, 4, 45}
\definecolor{successgreen}{RGB}{4, 150, 4}
\definecolor{infoblue}{RGB}{0, 110, 255}
\definecolor{error_bg}{RGB}{255, 235, 238}
\definecolor{error_frame}{RGB}{220, 53, 69}
\definecolor{success_bg}{RGB}{232, 245, 233}
\definecolor{success_frame}{RGB}{40, 167, 69}
\definecolor{info_bg}{RGB}{232, 244, 253}
\definecolor{info_frame}{RGB}{23, 162, 184}

\newtcolorbox{logbox}[1]{
  colback=#1_bg,
  colframe=#1_frame,
  boxrule=0.5pt,      
  arc=0mm,
  left=0mm,           
  right=0mm,
  top=0mm,            
  bottom=0mm,
  fontupper=\ttfamily\scriptsize,
  before skip=1pt,
  after skip=2pt,
  leftrule=0pt,       
  rightrule=0pt       
}

\lstdefinestyle{log_style}{
    basicstyle=\ttfamily\scriptsize,
    backgroundcolor=\color{codebg},
    breaklines=true,
    frame=tb,
    framerule=0.5pt,
    columns=fullflexible,
    keepspaces=true,
    keywords=[1]{[ERROR]},
    keywords=[2]{[SUCCESS]},
    keywords=[3]{[INFO]},
    keywordstyle=[1]\color{errorred},
    keywordstyle=[2]\color{successgreen},
    keywordstyle=[3]\color{infoblue},
}

\makeatletter
\newcommand{\TODO}[1]{%
 \bgroup
 \def\@tempa{#1}%
 \expandafter\textcolor\expandafter{red}{\@tempa}%
 \GenericWarning{}{LaTeX Warning: TODO: \@tempa}%
 \egroup
}
\makeatother

\makeatletter
\newcommand{\NOTE}[1]{%
 \bgroup
 \def\@tempa{#1}%
 \expandafter\textcolor\expandafter{blue}{\@tempa}%
 \GenericWarning{}{LaTeX Warning: NOTE: \@tempa}%
 \egroup
}
\makeatother

\newcommand{\tool}{\textsc{\textsc{Meta Self-Refining}}}

\def\BibTeX{{\rm B\kern-.05em{\sc i\kern-.025em b}\kern-.08em
    T\kern-.1667em\lower.7ex\hbox{E}\kern-.125emX}}
\begin{document}

\title{Repairing Language Model Pipelines by \tool\ Competing Constraints at Runtime}

\author{\IEEEauthorblockN{Mojtaba Eshghie\orcidlink{0000-0002-0069-0588}}
\IEEEauthorblockA{\textit{KTH Royal Institute of Technology} \\
Stockholm, Sweden}
}

\maketitle

\begin{abstract}
Language Model (LM) pipelines can dynamically refine their outputs against programmatic constraints. However, their  effectiveness collapses when faced with competing \emph{soft} constraints, leading to inefficient backtracking loops where satisfying one constraint violates another. We introduce \tool\, a framework that equips LM pipelines with a meta-corrective layer to repair these competitions at runtime/inference-time. Our approach monitors the pipeline's execution history to detect oscillatory failures. Upon detection, it invokes a meta-repairer LM that analyzes the holistic state of the backtracking attempts and synthesizes a strategic instruction to balance the competing requirements. This self-repair instruction guides the original LM out of a failing refining loop towards a successful output. 
Our results show \tool\ can successfully repair these loops, leading to more efficient LM programs.
\end{abstract}


\section{Introduction}\label{sec:intro}

Language-model (LM) applications are increasingly expressed not as single prompts but as {pipelines} of modular LM calls that can reason, retrieve, and even invoke external tools~\cite{DSPy,DSPyAssertions}.  DSPy~\cite{DSPy} realises this paradigm by letting developers describe each call with a natural-language {signature} (for example, ``question~$\!\rightarrow\!$~answer''), implement that signature inside a parameterised \texttt{Module}, and check the module's output against \texttt{assertions} (for hard constraints) or \texttt{suggestions} (for soft constraints). 
The system is allowed to \emph{self-refine} its responses whenever a constraint is violated by backtracking and retrying the violated constraints until a certain maximum limit is reached~\cite{DSPyAssertions}. 


Self-refinement leverages a feedback loop during inference to iteratively improve outputs. Unlike reinforcement learning~\cite{OpenAIRL} or supervised fine-tuning~\cite{FT_LMs_ZeroShotLearners}, it operates entirely at runtime without requiring training~\cite{SelfRefine}.

\emph{Constraints can collide}.  A \texttt{suggestion} (lines 4 and 5 in \Cref{fig:tweet}) expresses a preference rather than an absolute requirement—the response is accepted after the maximum number of allowed retries, even if it continues to violate the suggestion during the runtime of the LM pipeline. Two suggestions may be individually reasonable yet difficult to satisfy together.  Consider the LM pipeline program in \Cref{fig:tweet}.  A valid response must propose a tweet (i) shorter than 100 characters and (ii) including certain keywords.  When the LM shortens a response to comply with the length constraint it often trims the sentence by excluding some of important keywords.  The naive back-tracking loop therefore oscillates, a \emph{ping-pong} failure that wastes the LM-call budget without producing an optimally compliant answer (as in \Cref{fig:results}). These failures might be especially susceptible in case of soft constraints (\texttt{suggestions}) as they only recommend behavior, and the pipeline never throws a hard error; instead it keeps retrying until the allotted budget expires. While developers can predict some of these competing constraints and use more directive error messages, it is often impractical to anticipate and manually script solutions for every potential negative interaction between complex constraints. The challenge calls for a dynamic, runtime approach rather than relying solely on developer foresight.

Recent work suggests that even a weaker, secondary observing LM can detect flaws in the reasoning of a frontier model~\cite{MonitoringReasoningModels}. The capability of detecting competing soft constraints (\texttt{suggestions}) through another LM or using heuristics on the stateful execution traces of modules enables the automated repair of LM pipelines affected by suboptimal outputs and costly ping-pong backtracking~\cite{APRSurvey}.

\setlength{\abovecaptionskip}{-3pt}
\begin{figure}[t]
\begin{lstlisting}[basicstyle=\fontsize{8}{9}\ttfamily,numberstyle=\footnotesize,
                   numbers=left,
                   numbersep=-2pt]
 class TweetSummarizer(dspy.Module):
  def forward(self, source_text, keywords):
   tweet = self.generate_tweet(source_text=source_text).tweet
   dspy.Suggest(contains_keywords(tweet, keywords), f"Tweet must include: {keywords}")
   dspy.Suggest(len(tweet) < 100, "Tweet must be < 100 characters.")
   return tweet
\end{lstlisting}
\caption{A LM pipeline program written in DSPy~\cite{DSPy} with soft competing constraints (\texttt{suggestions}) implementing a tweet summarization task.}
\label{fig:tweet}
\end{figure}

We propose \tool\footnote{\href{https://github.com/mojtaba-eshghie/Meta-Self-Refining}{https://github.com/mojtaba-eshghie/Meta-Self-Refining}}, a state-aware back-tracking scheme that detects these oscillations (\emph{ping-pong} failures), invokes a meta-repairer LM to synthesize a single instruction to balance and direct the {competing suggestions}, and feeds this instruction back into the failing module. \tool\ acts as a runtime repair mechanism, making it distinct from and complementary to compile-time optimizers like MIPRO~\cite{OptimizingInstructionsDemosMultiStageLMPrograms}. While optimizers fine-tune a pipeline  
before runtime, our tool dynamically resolves conflicts between multiple competing constraints during inference.

\section{The Meta-Self-Refining Architecture}\label{sec:Meta-Self-Refining-arch}

To solve the challenge of competing suggestions, we propose \tool\ as an extension to the runtime of LM pipelines (as in DSPy~\cite{DSPy}) to reason about the pipeline's own corrective trajectory. The solution consists of a runtime and a compile-time patching strategy.

In the DSPy, a program (LM pipeline) has two main lifecycle phases: compile-time and runtime.
Compile-time is an optimization step where DSPy compiler takes the LM pipeline as a Python program, a training set, and a metric, and optimizes the pipeline's behavior~\cite{OptimizingInstructionsDemosMultiStageLMPrograms}. This is typically done by bootstrapping few-shot demonstrations or by fine-tuning the LM's weights for some modules~\cite{DSPy,OptimizingInstructionsDemosMultiStageLMPrograms}. The result is a new, optimized LM program prepared for execution. Runtime or inference-time, is when the program is executed on new inputs to perform its task. During this phase, dynamic behaviors like self-refinement through backtracking on violated constraints are triggered by programmatic constraints (lines 4 and 5 in \Cref{fig:tweet}). This allows LM pipeline to correct its outputs on-the-fly. Our work adds a meta-repair layer to this runtime refining.

\subsection{Runtime: State-Aware Meta-Corrective Backtracking}\label{sec:runtime-solution}
At its core, \tool\ is an enhanced backtracking loop that actives when simple self-refining loops in an LM pipeline fails during runtime or inference-time. 
The mechanism involves four steps:
\begin{enumerate}[leftmargin=1em, itemsep=0.25ex, topsep=0pt, partopsep=0pt]
    \item \textbf{Loop Detection:} The system detects a backtracking loop by identifying repeating failure patterns (e.g., A fails → B fails → A fails again) and triggers the meta-repair process.
    \item \textbf{Context Aggregation:} Upon detecting a loop, the system gathers a rich \emph{state snapshot}. Instead of using a single error message, this snapshot includes all active constraints and suggestions that apply to the failing module to provide a more holistic view of the issue.
    \item \textbf{Meta-Repair:} It then invokes a \emph{meta-repair} which is another LM call tasked with analyzing this complete state. The meta-repairer's goal is not just to fix one error, but to synthesize a new instruction that directs the original model on how to balance the competing requirements. For instance, it might generate guidance like: ``Ensure the answer includes the term $X$, even if it slightly increases length by including mini-sentences with the keywords.'' for the constraints in \Cref{fig:tweet}.
    \item \textbf{Informed Retry:} The original module is then retried using this new instruction, which provides the precise guidance needed to break the loop and produce an output that respects a \emph{trade-off between the constraints}.
\end{enumerate}

\subsection{Compile-Time: Meta-Refining Compilation}
The runtime solution (\Cref{sec:runtime-solution}) introduces a potential drift between the repaired pipeline used during execution and the compile-time-optimized LM program. 
Therefore, we propose integrating the the \tool\ output directly into the \emph{counterexample bootstrapping} process of DSPy~\cite{DSPyAssertions}.
\begin{enumerate}[leftmargin=1em, itemsep=0.25ex, topsep=0pt, partopsep=0pt]
    \item During the few-shot bootstrap phase~\cite{DSPy}, a teacher model generates examples to prepare the LM pipeline.
    \item {If} a teacher-generated output violates a set of competing constraints, the compiler invokes the same \tool\ to generate a dynamic feedback instruction.
    \item The entire trace including the initial failure, the invocation of the meta-repair procedure, the dynamically generated instruction, and the subsequent successful retry is packaged as a \emph{counter-example} used in the counter-example-based bootstrapping.
\end{enumerate}
The few-shot demonstrations will teach the pipeline to react to the same failure scenarios with competing constraints.

\begin{figure}[t]
\setlength{\abovecaptionskip}{3pt}
\begin{logbox}{error}
Attempt 1 Failed: Tweet must be very concise (under 100 characters). 
(Tweet: 'Generative Adversarial Networks (GANs), created by Ian Goodfellow, involve a generator and discriminator competing...')
\end{logbox}
\begin{logbox}{error}
Attempt 2 Failed: Tweet must include these keywords: ['GAN', 'generator', 'discriminator']. 
(Tweet: 'GANs: two neural networks compete--one creates, the other detects fake data...')
\end{logbox}
\begin{logbox}{error}
Attempt 3 Failed: Tweet must be very concise (under 100 characters). 
(Tweet: 'Generative Adversarial Networks (GANs) involve a generator creating data and a discriminator distinguishing real from fake...')
\end{logbox}

\begin{logbox}{info}
--- META-SELF-REFINING: PING-PONG LOOP DETECTED ---
Competition History: [('...long tweet...', 'Tweet must be very concise...'), ('...missing keyword tweet...', 'Tweet must include...')]
Synthesized Instruction: Create a concise tweet under 100 characters that includes the keywords 'GAN', 'generator', and 'discriminator'...
\end{logbox}

\begin{logbox}{success}
Success: GANs: generator creates data, discriminator detects fakes--adversaries in AI. \#AI \#GAN
\end{logbox}

\caption{Execution log of the \tool\ process for the example in \Cref{fig:tweet}.}
\label{fig:results}
\end{figure}








\section{Preliminary Results}\label{sec:results}

To validate the efficacy of \tool{}, we conducted a preliminary experiment based on a tweet summarizer program in \Cref{fig:tweet}. The objective of this program is to summarize a dense technical paragraph  
into a tweet, subject to two competing constraints (lines 4 and 5 in \Cref{fig:tweet}).
The experiment was configured using \texttt{GPT-4.1-nano} as the underlying LM for all roles: the base tweet generator, the teacher model during compilation, and the meta-repairer LM.

The execution trace, shown in \Cref{fig:results}, demonstrates a ping-pong failure mode. The pipeline's attempt 1 produces a summary that includes all keywords but violates the length constraint. In its first self-refinement step, the model overcorrects by shortening the tweet, which now satisfies the length constraint but omits the mandatory a keyword. The third attempt corrects this by re-inserting the keywords, but again violates the character limit, creating a \emph{(length $\rightarrow$ keyword $\rightarrow$ length)} failure pattern detected by the monitoring component of \tool. Next, \tool\ invokes the meta-repairer giving it the full context of the competing attempts, including the generated text and the associated error for each. Meta-repairer then synthesizes a new instruction that provides a strategy to resolve the competition loop.

This new instruction guides the base LM to a successful output that artfully satisfies both constraints. Without \tool, the pipeline would have continued its futile backtracking for the maximum number of retries (e.g., five), terminating on a suboptimal output that violates at least one of the constraints. Our experiment shows that \tool\ not only preempts this wasteful process but also repairs the pipeline to produce a higher-quality compliant result.

\section{Conclusion}\label{sec:conclusion}
We introduced \tool, a framework that addresses the problem of competing soft constraints in LM programs during inference-time by implementing a meta refinement process. By detecting failed ping-pong backtracking loops and invoking a meta-repair LM to synthesize holistic, context-aware instructions, our approach enables LM pipelines to balance competing constraints at runtime. Future work will explore the application of this framework to more complex, multi-agent systems where competing goals are more prevalent.

\bibliographystyle{ieeetr}
\bibliography{refs}

\begin{thebibliography}{1}

\bibitem{DSPy}
O.~Khattab, A.~Singhvi, P.~Maheshwari, Z.~Zhang, K.~Santhanam, S.~V.~A. Vardhamanan, S.~Haq, A.~Sharma, T.~T. Joshi, H.~Moazam, H.~Miller, M.~Zaharia, and C.~Potts, ``{DSPy: Compiling Declarative Language Model Calls into Self‑Improving Pipelines},'' in {\em {NeurIPS 2023 Workshop on Robustness of Zero/Few‑Shot Learning in Foundation Models (R0‑FoMo)}}, 2023.

\bibitem{DSPyAssertions}
A.~Singhvi, M.~Shetty, S.~Tan, C.~Potts, K.~Sen, M.~Zaharia, and O.~Khattab, ``{DSPy Assertions: Computational Constraints for Self-Refining Language Model Pipelines},'' 2024.

\bibitem{OpenAIRL}
D.~M. Ziegler, N.~Stiennon, J.~Wu, T.~B. Brown, A.~Radford, D.~Amodei, P.~Christiano, and G.~Irving, ``Fine-tuning language models from human preferences,'' 2020.

\bibitem{FT_LMs_ZeroShotLearners}
J.~Wei, M.~Bosma, V.~Y. Zhao, K.~Guu, A.~W. Yu, B.~Lester, N.~Du, A.~M. Dai, and Q.~V. Le, ``Finetuned language models are zero-shot learners,'' 2022.

\bibitem{SelfRefine}
A.~Madaan, N.~Tandon, P.~Gupta, S.~Hallinan, L.~Gao, S.~Wiegreffe, U.~Alon, N.~Dziri, S.~Prabhumoye, Y.~Yang, S.~Gupta, B.~P. Majumder, K.~Hermann, S.~Welleck, A.~Yazdanbakhsh, and P.~Clark, ``{SELF-REFINE: Iterative Refinement with Self-Feedback},'' in {\em Proceedings of the 37th International Conference on Neural Information Processing Systems}, NIPS '23, (Red Hook, NY, USA), Curran Associates Inc., 2023.

\bibitem{MonitoringReasoningModels}
B.~Baker, J.~Huizinga, L.~Gao, Z.~Dou, M.~Y. Guan, A.~Madry, W.~Zaremba, J.~Pachocki, and D.~Farhi, ``{Monitoring Reasoning Models for Misbehavior and the Risks of Promoting Obfuscation},'' 2025.

\bibitem{APRSurvey}
Q.~Zhang, C.~Fang, Y.~Ma, W.~Sun, and Z.~Chen, ``A survey of learning-based automated program repair,'' {\em ACM Trans. Softw. Eng. Methodol.}, vol.~33, Dec. 2023.

\bibitem{OptimizingInstructionsDemosMultiStageLMPrograms}
K.~Opsahl-Ong, M.~J. Ryan, J.~Purtell, D.~Broman, C.~Potts, M.~Zaharia, and O.~Khattab, ``{Optimizing Instructions and Demonstrations for Multi-Stage Language Model Programs},'' 2024.

\end{thebibliography}

\end{document}